\documentclass{ifacconf}

\usepackage{graphicx}      
\usepackage{natbib}        

\usepackage{comment}   
\usepackage{times}

\usepackage{graphics} 
\usepackage{epsfig} 
\usepackage{amsmath}
\usepackage{amssymb} 
\usepackage{theorem} 

\usepackage{bm}
\usepackage{ascmac}
\usepackage{subfigure}
\usepackage{color}
\usepackage{multicol}

\newcommand{\R}{{\mathbb{R}}}

\renewcommand{\S}{{\mathsf{\Sigma}}}

\newcommand{\dx} {{\rm d}x}

\newtheorem{definition}{Definition}

\newtheorem{theorem}{Theorem}

\newtheorem{remark}{Remark}
\begin{document}
\begin{frontmatter}

\title{Dissipativity of nonlinear ODE model of distribution 
 voltage profile
~\thanksref{footnoteinfo1}\thanksref{footnoteinfo2}} 

\thanks[footnoteinfo1]{
This work has been submitted to IFAC World Congress 2023 for possible publication. 
}
\thanks[footnoteinfo2]{
This work was supported by 
JSPS KAKENHI Grant Number 20K04552, ``Multi-scale robust control system design for next-generation power grids based on multi-dimensional systems and dissipativity''.
}

\author[First]{Chiaki Kojima} 
\author[Second]{Yuya Muto} 
\author[Third]{Yoshihiko Susuki}

\address[First]{Department of Electrical and Electronic Engineering, Faculty of Engineering, Toyama Prefectural University, 5180 Kurokawa, Imizu, Toyama, 939-0398, Japan (e-mail: chiaki@pu-toyama.ac.jp).}
\address[Second]{Department of Electrical and Computer Engineering, Graduate School of Engineering, Toyama Prefectural University, 
5180 Kurokawa, Imizu, Toyama, 939-0398, Japan (e-mail: u255019@st.pu-toyama.ac.jp)}
\address[Third]{Department of Electrical Engineering, Graduate School of Engineering, Kyoto University, Katsura, Nishikyo-ku, Kyoto, 615-8510, Japan\\
(e-mail: susuki.yoshihiko.5c@kyoto-u.ac.jp)}

\begin{abstract}        
In this paper, we consider a power distribution system consisting of a straight feeder line. A nonlinear ordinary differential equation (ODE) model is used to describe the voltage distribution profile over the feeder line. At first, we show the dissipativity of the subsystems corresponding to active and reactive powers. We also show that the dissipation rates of these subsystem coincide with the distribution loss given by a square of current amplitudes. Moreover, the entire distribution system is decomposed into two subsystems corresponding to voltage amplitude and phase. As a main result, we prove the dissipativity of these subsystems based on the decomposition. As a physical interpretation of these results, we clarify that the phenomena related to the gradients of the voltage amplitude and phase are induced in a typical power distribution system from the dissipation equalities. Finally, we discuss a reduction of distribution losses by injecting a linear combination of the active and reactive powers as a control input based on the dissipation rate of the subsystem corresponding to voltage amplitude. 
\end{abstract}


\end{frontmatter}


\section{Introduction}
\label{sec:intr}

A distribution system is a part of electric power systems that supply individual consumers with power that is transmitted over distribution lines (See \cite{mac:powe1}). In recent years, sustainable operation is required to deal with the increased uncertainty caused by the active introduction of distributed power sources, e.g. photovoltaic power generation, and electric vehicles (EVs) from a viewpoint of microgrids. To achieve this type of operation, many theoretical studies have been conducted to provide plug and play property to the power distribution system. In order to have this property, it is useful to give passivity or dissipativity (\cite{wil:leas1}, \cite{wil:diss1}, \cite{wil:diss2}) to the distribution system (\cite{arc:pass1}, \cite{qu:modu1}). 


The nonlinear ordinary differential equation (ODE) model~(See \cite{che:volt1}, \cite{tad:asym1}) is one of the mathematical models that describe the spatial voltage distribution of a power distribution system~(\cite{tad:asym1}) clarified the effect of the connection location of EVs (electric vehicles) on the voltage of power distribution lines. However, despite the importance of the plug and play properties discussed above, the discussion of dissipativity has been not been sufficiently addressed for the nonlinear ODE model. On the other hand, from the perspective of dissipation theory, there have been many studies on one-dimensional systems in which time is an independent variable. However, there have not been sufficiently studied for one-dimensional systems that do not necessarily require causality, such as space to the best of authors' knowledge.

Following the problems pointed out in the above paragraphs, this paper derives the dissipation equality for the nonlinear ODE model. Moreover, by extracting the dissipation equality corresponding to the power loss in this model, we aim to obtain new insights into the minimization of distribution losses. In particular, algebraic approaches based on power flow computations have been mainly adopted in the traditional framework. However, difficulties existed in the computation of optimal solutions due to the requirements to solve nonconvex optimization problems (See \cite{gan:exac1}). On the other hand, the framework of this paper is an approach based on dynamical systems. Therefore, by utilizing the knowledge from the system and control theory, it is possible to find the possibility of developing a control system design.

The outline of this paper is described as follows. In Section~\ref{sec:ode}, we formulate the nonlinear ODE model for power distribution systems, and show the dissipativity of the subsystems corresponding to the active and reactive powers. In Section~\ref{sec:deco}, the entire power distribution system is decomposed to the subsystems corresponding to the voltage amplitude and phase. As a main result, we prove the dissipativity of the subsystems, and derive dissipation equalities for each subsystem. This yields a physical interpretation related to the phenomena which are observed in a typical power distribution system. Finally, we provide a discussion for a reduction of distribution losses based on the dissipation rate of the subsystem corresponding to the voltage amplitude. 

\section{Dissipativity of nonlinear ODE model}
\label{sec:ode}

In this section, we first introduce a nonlinear ordinary differential equation (ODE) model (\cite{che:volt1}, \cite{tad:asym1}) for the voltage distribution profile of a power distribution system. We decompose the entire system into the subsystems corresponding to the active and reactive powers. Then, we prove the dissipativity of these subsystems. We also show a physical interpretation of the dissipation rates derived in the above. See Appendix~\ref{sec:prel} for the dissipation theory for one-dimensional nonlinear systems with independent variable which is not necessarily causal. 

\subsection{Nonlinear ODE model}
\label{sec:nonl}

\begin{figure}[htb]\centering
\includegraphics[width=0.95\columnwidth]{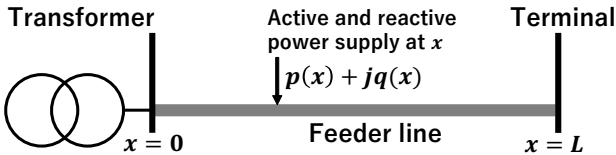}
\caption{Power distribution system in the straight feeder line considered in this paper.}
\label{zentai}
\end{figure}

Throughout this paper, we consider a power distribution system where the feeder line is extended in a straight line from the transformer to the terminal as shown in Fig.~\ref{zentai}. Note that electrical loads and electric power generations are typically connected at discrete locations on the feeder line. The electrical load is supposed to consume power from the feeder line, and the electric power generation, e.g. PV (photovoltaic) power generations are assumed to supply power to the line. For EVs, it is supposed not only to consume but also to supply power electricity. We assume that no electrical load is connected at the terminal of the feeder line. 

Mathematically, the position variable on the feeder line is defined as 
$x\mathrm{\, [m]}\in\mathbb{R}$. We suppose that $x=0$~[m] for the 
transformer (starting point) and $x=L$~[m], for the terminal of the 
feeder line, where $L>0$ is the length of the feeder line. Let 
$p(x)\mathrm{\,{{[W/km]}}}\in\mathbb{R}$ and 
$q(x)\mathrm{\,[Var/km]}\in\mathbb{R}$ be the active and reactive powers 
at position $x$ over the feeder line, respectively. If $p(x)>0$ holds, 
$p(x)$ denotes the active power supply flowing to the feeder line at $x$. 
On the other hand, $p(x)<0$ holds, $p(x)$ is the active power 
consumption flowing from the line at $x$. The same physical meaning 
holds true for the reactive power $q(x)$. As we have assumed that no 
electrical load is connected at the terminal of the feeder line, 
{{the 
following boundary conditions also hold for active and reactive powers at the terminal:
\begin{align}
p(L)=0, \ q(L)=0.
\label{p05}
\end{align}}}

Under the above setting on the feeder line, the spatial variation of the distribution voltage profile on the line is described by the \emph{nonlinear ODE model}~(See \cite{che:volt1}, \cite{tad:asym1}), 
which is the following set of the ODEs~\eqref{p01}-\eqref{p04}: 
\begin{align}
\label{p01}
\frac{{{\rm d}}\theta(x)}{{\rm d} x}&=-\frac{s(x)}{v(x)^2},\\
\label{p02}
\frac{{{\rm d}} v(x)}{{{\rm d}} x}&=w(x),\\
\label{p03}
\frac{{{\rm d}} s(x)}{{{\rm d}} x}&=\frac{bp(x)-gq(x)}{g^2+b^2},\\
\label{p04}
\frac{{{\rm d}} w(x)}{{{\rm d}} x}&=\frac{s(x)^2}{v(x)^3}-\frac{gp(x)+bq(x)}{\left(g^2+b^2\right)v(x)}, 
\end{align}
where $\theta(x)\mathrm{\,[rad]}\in\mathbb{R}$, 
$v(x)\mathrm{\,[V]}\in\mathbb{R}$, 
$s(x)\mathrm{\,[V^2/km]}\in\mathbb{R}$, and 
$w(x)\mathrm{\,[V/km]}\in\mathbb{R}$ represent the voltage phase, 
voltage amplitude, supplemental variable, and {{voltage gradient}} at 
position $x$ of the feeder line, respectively. 
In \eqref{p03} and \eqref{p04}, $g\mathrm{\,{{[S/km]}}}\in\mathbb{R}$ is 
the conductance per unit length at position $x$, 
which are assumed to be constant and positive, i.e. $g>0$. 
Moreover, $b\mathrm{\,{{[S/km]}}}\in\mathbb{R}$ is the susceptance per unit length at position $x$, which are assumed to be constant. 
Note that the susceptance $b$ can take both positive and negative values 
different from the conductance $g$. 
The boundary condition for the set of ODEs~\eqref{p01}-\eqref{p04} is given by the following 
equalities~(See \cite{che:volt1}, \cite{tad:asym1}) in per unit (p.u.) value: 
\begin{align}
\theta(0) = 0, \ v(0) = 1, \ s(L) = 0, \ w(L) = 0.
\label{p05}
\end{align}
In \eqref{p05}, the first and second equalities imply that we use the 
voltage amplitude $v(0)$ and phase $\theta(0)$ at the transformer as the 
reference values. 
From the assumption that no electrical load is connected at the terminal, the third and fourth equalities are imposed as the boundary condition.

\begin{figure}[htb]
\centering
\includegraphics[width=0.65\columnwidth]{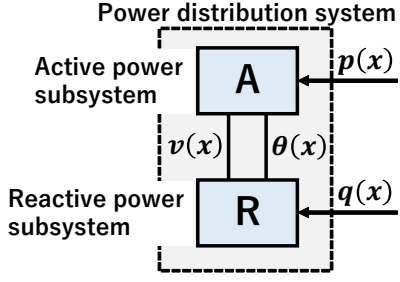}
\caption{Block diagram of the active and reactive power subsystems}
\label{block}
\end{figure}

By gathering the ODEs~\eqref{p01}-\eqref{p04}, we can be prove that the active power $p(x)$ and reactive power $q(x)$ satisfy the following ODEs in terms of the voltage amplitude $v(x)$ and phase $\theta(x)$: 
\begin{align}
p(x)&+b\left(2v(x)\frac{{{\rm d}} v(x)}{{{\rm d}} x}\frac{{{\rm d}}\theta(x)}{{\rm d} x}
+v(x)^2\frac{{{\rm d}}^2\theta(x)}{{\rm d} x^2}\right)
\nonumber \\
& +g\left\{v(x)\frac{{{\rm d}^2} v(x)}{{{\rm d}} x^2}
-v(x)^2\left(\frac{{{\rm d}}\theta(x)}{{\rm d} x}\right)^{{2}}\right\}=0,
\label{p06}\\
q(x)&+b\left\{v(x)\frac{{{\rm d}^2} v(x)}{{{\rm d}} x^2}
-v(x)^2\left(\frac{{{\rm d}}\theta(x)}{{\rm d} x}\right)^2\right\}
\nonumber \\
& -g\left(2v(x)\frac{{{\rm d}} v(x)}{{{\rm d}} x}\frac{{{\rm d}}\theta(x)}{{\rm d} x}
+v(x)^2\frac{{{\rm d}}^2\theta(x)}{{\rm d} x^2}\right)=0.
\label{p07}
\end{align}
In this paper, the systems described by the ODEs~\eqref{p06} and \eqref{p07} are called an \emph{active power subsystem} $\mathsf{A}$ and a \emph{reactive power system} $\mathsf{R}$, respectively. From a system and control viewpoint, the voltage amplitude $v(x)$ and phase $\theta(x)$ can be regarded as the states (and become the outputs simultaneously) of each subsystem. Also, the active power $p(x)$ and reactive power $q(x)$ can be regarded as inputs to each subsystem. The block diagram of these subsystems is shown in Fig.~\ref{block}. Note that there is no clear input-output relationship between the two subsystems via $v(x)$ and $\theta(x)$. Therefore, we follow the framework of the behavioral system theory (See \cite{wil:para1}) that enables us to treat such cases. For this reason, the tips of the arrows are removed in the diagram. 

\subsection{Dissipativity of active and reactive power subsystems}
\label{sec:equa}

In terms of a complex-valued power, 
the ODEs~\eqref{p06} and \eqref{p07} of the active and reactive power subsystems can be combined into a single ODE 
\begin{align}
& \left(p(x)+jq(x)\right) 
\nonumber \\
& + \left(b-jg\right)\left(2v(x)\frac{{{\rm d}} v(x)}{{{\rm d}} x}\frac{{{\rm d}}\theta(x)}{{\rm d} x}
+v(x)^2\frac{{{\rm d}}^2\theta(x)}{{\rm d} x^2}\right)
\nonumber \\
& + \left(g+jb\right)\left\{v(x)\frac{{{\rm d}^2} v(x)}{{{\rm d}} x^2}
-v(x)^2\left(\frac{{{\rm d}}\theta(x)}{{\rm d} x}\right)^{{2}}\right\}=0.
\label{d01}
\end{align}
From a viewpoint of dissipation theory, we can also regard the active 
power $p(x)$ and reactive power $q(x)$ as the supply rates at position 
$x$ for the subsystems $\mathsf{A}$ and $\mathsf{R}$, respectively. 
The reason is that these power supplies electric powers to the considered power 
distribution system. Thus, we can investigate dissipativity for each subsystem.


To examine the dissipativity for the subsystems $\mathsf{A}$ and $\mathsf{R}$, 
we define the function 
$\Psi_{\rm b}: \R \times \R \rightarrow \R$ by 
\begin{align}
\Psi_{\rm b}(v(x),\theta(x)):=-v(x)^2\frac{{\rm d}\theta(x)}{{\rm d}x}.
\label{d02}
\end{align}
We can prove that this function  coincides  with the supplemental 
variable $s(x)$ in the nonlinear ODE model~\eqref{p01}-\eqref{p04} by a 
simple computation. 
The gradient of $\Psi_{\rm b}(v(x),\theta(x))$ satisfies the equality
\begin{align}
2v(x)\frac{{{\rm d}} v(x)}{{{\rm d}} x}\frac{{{\rm d}}\theta(x)}{{\rm d} x}
+v(x)^2\frac{{{\rm d}}^2\theta(x)}{{\rm d} x^2}
=-\frac{{\rm d}\Psi_{\rm b}(v(x),\theta(x))}{{\rm d}x}.
\label{d03}
\end{align}
In addition to $\Psi_{\rm b}(v(x),\theta(x))$, we define the function $\Psi_{\rm g}: \R \rightarrow \R$ by 
\begin{align}
\Psi_{\rm g}(v(x)):=-v(x)\frac{{\rm d}v(x)}{{\rm d}x}.
\label{d04}
\end{align}
By computing the gradient of this function, 
the function satisfies the equality 
\begin{align}
v(x)\frac{{\rm d}^2v(x)}{{\rm d}x^2}
=-\frac{{\rm d}\Psi_{\rm g}(v(x))}{{\rm d}x}
-\left(\frac{{\rm d}v(x)}{{\rm d}x}\right)^2.
\label{d05}
\end{align}


Substituting \eqref{d03} and \eqref{d05} into the first and second 
terms of the left-hand side of \eqref{d01}, respectively, we have the equality
\begin{align*}
& \left(p(x)+jq(x)\right) 
\nonumber \\
& - \left(b-jg\right)\frac{{\rm d}\Psi_{\rm b}(v(x),\theta(x))}{{\rm d}x}
- \left(g+jb\right)\frac{{\rm d}\Psi_{\rm g}(v(x))}{{\rm d}x}
\nonumber \\
& -\left(g+jb\right)\left\{\left(\frac{{\rm d}v(x)}{{\rm d}x}\right)^2
+v(x)^2\left(\frac{{{\rm d}}\theta(x)}{{\rm d} x}\right)^{{2}}\right\}=0.
\end{align*}
By solving for the gradient of the linear combinations of the two functions 
in \eqref{d02} and \eqref{d04}, we have
\begin{align}
& \frac{{\rm d}}{{\rm d}x}\left(b\Psi_{\rm b}(v(x),\theta(x))+g\Psi_{\rm g}(v(x))\right)
\nonumber \\
&+j\frac{{\rm d}}{{\rm d}x}\left(
b\Psi_{\rm g}(v(x))-g\Psi_{\rm b}(v(x),\theta(x))\right)
\nonumber \\ 
& = \left(p(x)+jq(x)\right)-\left(g+jb\right)\Delta(v(x),\theta(x)),
\label{d08}
\end{align}
where $\Delta: \R \times \R \rightarrow \R$ is the quadratic function 
defined by 
\begin{align}
\Delta(v(x),\theta(x)):=
\left(\frac{{\rm d}v(x)}{{\rm d}x}\right)^2
+v(x)^2\left(\frac{{{\rm d}}\theta(x)}{{\rm d} x}\right)^{{2}}.
\label{d07}
\end{align}
Note that it is clear that $\Delta(v(x),\theta(x))$ is a positive definite 
function from the above equation. The real part of \eqref{d08} is described by the following equality: 
\begin{align}
\frac{{\rm d}}{{\rm d}x}&\left(b\Psi_{\rm b}(v(x),\theta(x))+g\Psi_{\rm g}(v(x))\right)
\nonumber \\
& = p(x)-g\Delta(v(x),\theta(x)).
\label{d09}
\end{align}
From the imaginary part of \eqref{d08}, 
we can also derive the equality  
\begin{align}
\frac{{\rm d}}{{\rm d}x}&\left(b\Psi_{\rm g}(v(x))-g\Psi_{\rm b}(v(x),\theta(x))\right)
\nonumber \\ 
& = q(x)-b\Delta(v(x),\theta(x)).
\label{d10}
\end{align}
Based on these equalities, we have the following theorem for the dissipativity of the active and reactive power subsystems in the sense of dissipation equality. The proof follows by applying Definition~\ref{df3} to the equalities \eqref{d09} and \eqref{d10}.
\begin{theorem}
\label{tm1}
Consider the active power subsystem~$\mathsf{A}$ in \eqref{p06} and reactive power 
system~$\mathsf{R}$ in \eqref{p07}. 
Then, we have the following statements~(i) and (ii). 
\begin{itemize}
\item[(i)]
{{The active power subsystem~$\mathsf{A}$ is dissipative with respect to 
	  the supply rate $p(x)$. 
Moreover,}} 
the functions $b\Psi_{\rm b}(v(x),\theta(x))+g\Psi_{\rm 
	  g}(v(x))$ and $g\Delta(v(x),\theta(x))$ are the {{flux}} 
and dissipation rate, respectively, with respect to  the supply rate $p(x)$. 
\item[(ii)]
{{If $b \geq 0$ holds, 
the reactive power subsystem~$\mathsf{R}$ is dissipative with respect to 
	  the supply rate $q(x)$. 
Moreover}}, the functions $b\Psi_{\rm g}(v(x))-g\Psi_{\rm b}(v(x),\theta(x))$ and $b\Delta(v(x),\theta(x))$ are the {{flux}} and dissipation rate, respectively, with respect to the supply rate $q(x)$. 
If $b<0$ holds, $\mathsf{R}$ is not dissipative with respect to  the supply rate $q(x)$.
\end{itemize}
\end{theorem}

In Theorem~1, we have shown that the functions $g\Delta(v(x),\theta(x))$ and $b\Delta(v(x),\theta(x))$ correspond to the dissipation rates in the active and reactive power subsystems. In the next subsection, we verify that these dissipation rates are consistent with the well-known distribution loss in power systems.

\subsection{Physical interpretation of dissipation rates}
\label{sec:delt}

In this subsection, we discuss an interpretation of the function 
$\Delta(v(x),\theta(x))$ 
corresponding to the dissipation rates in both active and reactive subsystems.

We consider the phasor representation $\dot{v}(x)=v(x)e^{j\theta(x)} \in 
\mathbb{C}$ 
of the voltage at position $x$, 
where the symbol ``$\cdot$'' over a complex variable stands for the phasor 
representation of the variable, not time derivative of the variable. 
The gradient of this phasor representation and its complex conjugate are 
computed by the following equations: 
\begin{align*}
\frac{{\rm d}\dot{v}(x)}{{\rm d}x} 
&= \frac{{\rm d}}{{\rm d}x}\left(v(x)e^{j\theta(x)}\right)\\
&= \frac{{\rm d}v(x)}{{\rm d}x}e^{j\theta(x)}
+ v(x)\frac{{\rm d}}{{\rm d}x}e^{j\theta(x)}\\
&= \frac{{\rm d}v(x)}{{\rm d}x}e^{j\theta(x)}
+ jv(x)e^{j\theta(x)}\frac{{\rm d}\theta(x)}{{\rm d}x}, \\ 
\left(\frac{{\rm d}\dot{v}(x)}{{\rm d}x}\right)^* 
&= \frac{{\rm d}}{{\rm d}x}\left(v(x)e^{-j\theta(x)}\right)\\
&= \frac{{\rm d}v(x)}{{\rm d}x}e^{-j\theta(x)}
- jv(x)e^{-j\theta(x)}\frac{{\rm d}\theta(x)}{{\rm d}x},
\end{align*}
where the symbol ``$*$'' denotes the complex conjugate. 
Then, we can prove that 
the product of the gradient and the complex conjugate coincides with $\Delta(v(x),\theta(x))$ as 
follows: 
\begin{align*}
& \hspace*{-6mm}
\left(\frac{{\rm d}\dot{v}(x)}{{\rm d}x}\right)^*
\left(\frac{{\rm d}\dot{v}(x)}{{\rm d}x}\right)\\
&= \left(\frac{{\rm d}v(x)}{{\rm d}x}e^{-j\theta(x)}
- jv(x)e^{-j\theta(x)}\frac{{\rm d}\theta(x)}{{\rm d}x}\right)\\
& \quad \cdot \left(\frac{{\rm d}v(x)}{{\rm d}x}e^{j\theta(x)}+ jv(x)e^{j\theta(x)}\frac{{\rm d}\theta(x)}{{\rm d}x}\right)\\
&= \left(\frac{{\rm d}v(x)}{{\rm d}x}\right)^2+v(x)^2\left(\frac{{\rm d}\theta(x)}{{\rm d}x}\right)^2\\
&= \Delta(v(x),\theta(x)).
\end{align*}

We also see that 
the gradient $\frac{{\rm d}\dot{v}(x)}{{\rm 
d}x}$ of the voltage can be described by the phasor representation of the 
current $\dot{I}(x) \in \mathbb{C}$ at position $x$ as follows: 
\begin{align*}
\frac{{\rm d}\dot{v}(x)}{{\rm d}x} 
&= \lim_{\delta x \rightarrow 0}\frac{v(x+\delta x)e^{j\theta(x+\delta 
 x)}-v(x)e^{j\theta(x)}}{\delta x}\\
& = \left(R+jX\right)\lim_{\delta x \rightarrow 0}\frac{v(x+\delta x)e^{j\theta(x+\delta 
 x)}-v(x)e^{j\theta(x)}}{(R+jX)\delta x}\\
&= \left(R+jX\right)\dot{I}(x),
\end{align*}
where $R>0$ and $X \in \mathbb{R}$ are the resistance and reactance of 
the feeder line per unit length, respectively. 
Note that these constants satisfy the equality $\left(R+jX\right)^{-1}=g-jb$. 
By computing the second term corresponding to the dissipation rate 
on the right-hand side of \eqref{d08}, we have 
\begin{align}
& \left(g+jb\right)\Delta(v(x),\theta(x))
\nonumber \\
&\quad = \left(g+jb\right)\left\{\left(R+jX\right)\dot{I}(x)\right\}^*\left\{\left(R+jX\right)\dot{I}(x)\right\}
\nonumber \\
&\quad = \left(R+jX\right)I(x)^2,
\label{d15}
\end{align}
where $I(x):=\left|\dot{I}(x)\right|$ is the current amplitude of $\dot{I}(x)$. 
The real part $RI(x)^2$ of the right-hand side of \eqref{d15} 
satisfies the equality 
\begin{align*}
g\Delta(v(x),\theta(x)) = RI(x)^2.
\end{align*}
Thus, we can verify that this real part is consistent with the well-known fact 
that the loss due to active power distribution is proportional to the square of 
the current $I(x)$~(See \cite{sug:anal1}). 
We can also prove that, if $b>0$ holds, 
the distribution loss of the reactive power can be represented by the 
square of the current as follows: 
\begin{align*}
b\Delta(v(x),\theta(x)) = XI(x)^2.
\end{align*}

\section{Dissipativity of voltage and phase subsystems}
\label{sec:deco}

In this section, the entire distribution system considered in the previous section is transformed to the subsystems corresponding to the voltage amplitude and phase. As a main result, we show the dissipativity of these subsystems. Furthermore, as a physical interpretation of these results, we show that the typical phenomena related to the gradients of the voltage amplitude and phase are induced from the dissipativity of these subsystems. Finally, we provide a discussion on the reduction of the distribution loss in the entire system. See Appendix~\ref{sec:prel} for the dissipation theory for one-dimensional nonlinear systems with possibly noncausal independent variable. 

\subsection{Decomposition into subsystems}
\label{sec:stor}

In this subsection, we consider the linear combinations of the active and 
reactive powers as new supply rates. Based on the combinations, the entire distribution system is decomposed into voltage subsystem $\mathsf{V}$ and phase subsystem $\mathsf{P}$, which are defined in the next paragraph. 

The ODEs~\eqref{p06} and \eqref{p07} can be gathered by defining  the appropriate vectors and constant matrices as follows: 
\begin{align}
& 
\begin{bmatrix}
p(x) \\ 
q(x) \\ 
\end{bmatrix}
+ 
\begin{bmatrix}
b & g \\ 
-g & b \\ 
\end{bmatrix}
{\small\begin{bmatrix}
2v(x)\frac{{{\rm d}} v(x)}{{{\rm d}} x}\frac{{{\rm d}}\theta(x)}{{\rm d} x}
+v(x)^2\frac{{{\rm d}}^2\theta(x)}{{\rm d} x^2} \\
v(x)\frac{{{\rm d}^2} v(x)}{{{\rm d}} x^2}
-v(x)^2\left(\frac{{{\rm d}}\theta(x)}{{\rm d} x}\right)^{{2}} \\ 
\end{bmatrix}}
= 
\begin{bmatrix}
0 \\ 
0 \\ 
\end{bmatrix}.
\label{e01}
\end{align}
Premultiplying \eqref{e01} by the orthogonal matrix $\frac{1}{g^2+b^2}
{\small\begin{bmatrix}
b & -g \\ 
g & b \\ 
\end{bmatrix}}$, we can derive the new equalities of the entire distribution 
system as 
\begin{align}
& 
\frac{1}{g^2+b^2}
\begin{bmatrix}
b & -g \\ 
g & b \\ 
\end{bmatrix}
\begin{bmatrix}
p(x) \\ 
q(x) \\ 
\end{bmatrix}
+ 
{\small\begin{bmatrix}
2v(x)\frac{{{\rm d}} v(x)}{{{\rm d}} x}\frac{{{\rm d}}\theta(x)}{{\rm d} x}
+v(x)^2\frac{{{\rm d}}^2\theta(x)}{{\rm d} x^2} \\
v(x)\frac{{{\rm d}^2} v(x)}{{{\rm d}} x^2}
-v(x)^2\left(\frac{{{\rm d}}\theta(x)}{{\rm d} x}\right)^{{2}} \\ 
\end{bmatrix}}
\nonumber \\
& = 
\begin{bmatrix}
0 \\ 
0 \\ 
\end{bmatrix}.
\label{e02}
\end{align}
Considering the first column of the above equation, we obtain the 
following ODE: 
\begin{align}
\frac{bp(x)-gq(x)}{g^2+b^2}
+ 2v(x)\frac{{{\rm d}} v(x)}{{{\rm d}} x}\frac{{{\rm d}}\theta(x)}{{\rm d} x}
+v(x)^2\frac{{{\rm d}}^2\theta(x)}{{\rm d} x^2}
= 0. 
\label{e03}
\end{align}
The subsystem described by this ODE is called the \emph{voltage subsystem} 
$\mathsf{V}$. 
Note that we can prove that the equalities~\eqref{e03} corresponds to the 
ODEs~\eqref{p03} which describes the derivative $w(x)$ of the gradient of the 
voltage amplitude $v(x)$. 
For this reason, we mention $\mathsf{V}$ as the voltage subsystem. 

In addition to the ODE \eqref{e03}, we have another ODE
\begin{align}
\hspace*{-1mm}
\frac{gp(x)+bq(x)}{g^2+b^2}
+ v(x)\frac{{{\rm d}^2} v(x)}{{{\rm d}} x^2}
-v(x)^2\left(\frac{{{\rm d}}\theta(x)}{{\rm d} x}\right)^{{2}} 
= 0
\label{e04}
\end{align}
from the second column of \eqref{e02}. We can be verify that the equality~\eqref{e04} corresponds to the ODE~\eqref{p04} which describes the derivative of the supplemental variable $s(x)$. Since $s(x)$ characterizes the derivative of the voltage phase $\theta(x)$, we call the subsystem the \emph{phase subsystem} $\mathsf{P}$. 

\begin{figure}[htb]
\centering
\includegraphics[width=0.9\columnwidth]{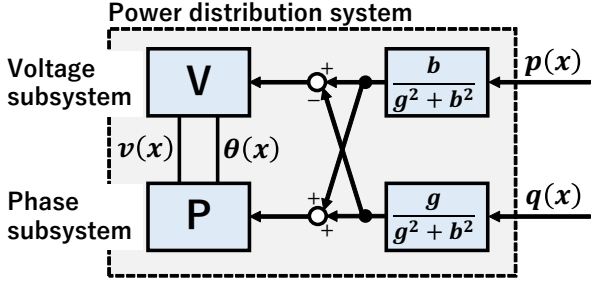}
\caption{Block diagram of the entire system consisting of the voltage and phase subsystems}
\label{diss}
\end{figure}

We provide the block diagram of the entire system consisting of the voltage and phase subsystems in Fig.~\ref{diss}. 

\subsection{Dissipativity of voltage and phase subsystems}

In this subsection, we show the dissipativity of the voltage 
and phase subsystems defined in the previous subsection. 
The result of this subsection can be regarded as a main result of this paper. 

From \eqref{d05} and \eqref{d07}, we have the equality
\begin{align*}
v(x)&\frac{{\rm d}^2v(x)}{{\rm d}x^2}
-v(x)^2\left(\frac{{{\rm d}}\theta(x)}{{\rm d} x}\right)^{{2}} \\
& =-\frac{{\rm d}\Psi_{\rm g}(v(x))}{{\rm d}x} - \Delta(v(x),\theta(x)).
\end{align*}
By substituting this equality and \eqref{d03} into \eqref{e02}, the equality~\eqref{e02} can be rewritten in terms of $\Psi_{\rm b}(v(x),\theta(x))$ and $\Psi_{\rm g}(v(x))$ as follows: 
\begin{align*}
\frac{1}{g^2+b^2}
\begin{bmatrix}
b & -g \\ 
g & b \\ 
\end{bmatrix}
\begin{bmatrix}
p(x) \\ 
q(x) \\ 
\end{bmatrix}
- \frac{\rm d}{{\rm d}x}
\begin{bmatrix}
\Psi_{\rm b}(v(x),\theta(x)) \\ 
\Psi_{\rm g}(v(x)) \\ 
\end{bmatrix}&\\
{{- 
\begin{bmatrix}
0 \\ 
\Delta(v(x),\theta(x)) \\ 
\end{bmatrix}=
\begin{bmatrix} 
0 \\ 
0 \\ 
\end{bmatrix}}}&.
\end{align*}
From the above equality, 
we have the following equalities on the gradient of $\Psi_{\rm g}(v(x))$ and $\Psi_{\rm b}(v(x),\theta(x))$: 
\begin{align}
\frac{{\rm d}\Psi_{\rm g}(v(x))}{{\rm d}x}
&=\frac{gp(x)+bq(x)}{g^2+b^2}
-\Delta(v(x),\theta(x)),
\label{e05}\\
\frac{{\rm d}\Psi_{\rm b}(v(x),\theta(x))}{{\rm d}x}
&=\frac{bp(x)-gq(x)}{g^2+b^2}.
\label{e06}
\end{align}

Based on the above computation, we have the following theorem which shows the dissipativity of the subsystems $\mathsf{V}$ and $\mathsf{P}$ in the sense of dissipation equality. This theorem can be regarded as a main result of this paper. The proof follows from the equalities~\eqref{e05}, \eqref{e06} and Definition~\ref{df3}. 
\begin{theorem}
\label{tm2}
Consider the voltage subsystem~$\mathsf{V}$ in \eqref{e05} and 
 phase subsystem~$\mathsf{P}$ in \eqref{e06}. 
Then, we have the following statements~(i) and (ii). 
\begin{itemize}
\item[(i)]
{{The voltage subsystem~$\mathsf{V}$ is dissipative with respect to 
	  the supply rate $\frac{gp(x)+bq(x)}{g^2+b^2}$. 
Moreover, 
the functions $\Psi_{\rm g}(v(x))$ and $\Delta(v(x),\theta(x))$ are the storage 
	  function and dissipation rate, respectively, with respect to 
	  the supply rate $\frac{gp(x)+bq(x)}{g^2+b^2}$. }}
\item[(ii)]
{{The phase subsystem~$\mathsf{P}$ is lossless with respect to 
	  the supply rate $\frac{bp(x)-gq(x)}{g^2+b^2}$.
Then, the functions $\Psi_{\rm 
	  b}(v(x),\theta(x))$ is the flux with respect to 
	  the supply rate $\frac{bp(x)-gq(x)}{g^2+b^2}$.}} 
\end{itemize}
\end{theorem}

In the next subsection, a physical interpretation of Theorem~\ref{tm2} 
is clarified in detail from a viewpoint of electrical phenomena related to the gradients of the voltage amplitude and phase in typical power distribution systems.

\subsection{Physical interpretation of main result}
\label{sec:phys2}

By computing the integrals of the supply rates in the dissipation 
equalities~\eqref{e05} and \eqref{e06}, 
and by substituting the boundary condition~\eqref{p05} into these integrals, 
we have the following equalities: 
\begin{align}
\hspace*{-2mm}\int_0^L\frac{gp(x)+bq(x)}{g^2+b^2}{\rm d}x - \frac{{\rm d}v(0)}{{\rm d}x} 
&= \int_0^L\Delta(v(x),\theta(x)){\rm d}x, 
\label{j02}\\ 
\hspace*{-2mm}\int_0^L\frac{bp(x)-gq(x)}{g^2+b^2}{\rm d}x
-\frac{{\rm d}\theta(0)}{{\rm d}x}&=0.
\label{j04}
\end{align}
The equalities~\eqref{j02} and \eqref{j04} are important because they 
characterize the relationship between the active and reactive powers, 
the transmission losses, and the gradients of the voltage amplitude and 
phase at the transformer. In particular, these  equalities correspond to 
the equalities obtained from the integral of the dissipation equalities of the voltage and phase subsystems, respectively. Each equality contains the gradients of the voltage amplitude and phase at the transformer. These facts are another reason that we call the subsystems described by the ODEs~\eqref{e05} and \eqref{e06} the voltage and phase subsystems, respectively.

In the remainder of this subsection, we address a conventional power 
distribution system consisting of only loads, and a current power distribution system including PVs and EVs as typical power distribution systems supposed in this paper. By applying Theorem~\ref{tm2} to this system, we clarify a physical interpretation of the theorem. 

\subsubsection{\ref{sec:phys2}.1 Conventional power distribution system: }

In the following, we consider a conventional power distribution system, 
and give a physical interpretation of the statement in Theorem~\ref{tm2}. 
Such distribution system is known to have the following properties on the susceptance and the active 
and reactive powers: 
\begin{itemize}
\item
The effect due to an inductance of the feeder line is dominant. 
This implies that the susceptance satisfies $b>0$ and $b \approx g$. 
\item
PVs and EVs are not installed to the feeder line. In addition, loads are connected to spatially discrete positions. This implies that the active and reactive powers have nonpositive values,  
i.e. $p(x) \leq  0$ and $q(x) \leq 0$ hold at any position $x$. 
\end{itemize}
Since the active and reactive powers are nonpositive, the integral of the supply rate of $\mathsf{V}$ is nonpositive in the left hand side of the \eqref{j02}:
\begin{align*}
\int_0^L \frac{gp(x)+bq(x)}{g^2+b^2} {\rm d}x \leq 0.
\end{align*}
For this reason, the dissipativity   is not satisfied in the sense of 
nonnegativity of the integral of the supply rate over the infinite 
lengths of the integral domain, which are considered in \cite{pil:loss1}. This is also due to the fact that we consider the space as a finite interval. On the other hand, since there is a power consumption of the connected load, a current flows from the feeder line to the load. This implies that distribution loss occurs over the line. Thus, the integral on the right-hand side of \eqref{j02} becomes nonnegative as follows: 
\begin{align*}
\int_0^L \Delta(v(x),\theta(x)){\rm d}x \geq 0.
\end{align*}
In such a case, for the equality~\eqref{j02} to hold, we have the inequality
\begin{align}
\frac{{\rm d}v(0)}{{\rm  d}x} \leq 0
\label{i06}
\end{align}
at the transformer ($x=0$). 
This inequality indicates the occurrence of a voltage drop on the feeder line (\cite{mac:powe1}). 

Moreover, considering \eqref{j04} for the phase subsystem $\mathsf{P}$, the sign of the phase gradient $\frac{{\rm d}\theta(0)}{{\rm d}x}$ at the transformer is indefinite depending on the values of the active and reactive powers from the equality
\begin{align*}
\frac{{\rm d}\theta(0)}{{\rm d}x}=\int_0^L \frac{bp(x)-gq(x)}{g^2+b^2}{\rm d}x.
\end{align*}
Combining the above equality with $q(x) \leq 0$, we see that the active and reactive powers satisfy the inequality
\begin{align*}
\frac{b}{g}\int_0^Lp(x){\rm d}x \leq \int_0^Lq(x){\rm d}x \leq 0
\end{align*}
if and only if the following inequality holds: 
\begin{align}
\frac{{\rm d}\theta(0)}{{\rm d}x} & \leq 0.
\label{i07}
\end{align}
The above inequality corresponds to the delay of the voltage phase which typically occurs in conventional power systems (\cite{mac:powe1}). 

The above series of the observation is consistent with typical phenomena observed in conventional power distribution systems, which shows a validity of our main result given in Theorem~\ref{tm2}. 

\subsubsection{\ref{sec:phys2}.2 Power distribution system with PVs and EVs:}

In a current power distribution system including PVs and EVs, the active and reactive powers at the position $x$, where the PVs or EVs are connected, may have positive values. This implies that $p(x)>0$ and $q(x)>0$ hold for the position $x$. In such a case, the following inequality typically holds in the first term on the left side of the \eqref{j02}: 
\begin{align*}
\int_0^L\frac{gp(x)+bq(x)}{g^2+b^2}{\rm d}x \geq 0.
\end{align*}
This implies that the integral of the supply rate of the voltage subsystem $\mathsf{V}$ possibly becomes nonnegative. Thus, contrary to the conventional power distribution systems, the dissipativity is guaranteed only by the integral of the supply rate. However, for some values of $p(x)$ and $q(x)$, the net power supply will exceed the distribution loss in the right side of the \eqref{j02}. Then, it is necessary that the inequality 
\begin{align}
\frac{{\rm d}v(0)}{{\rm d}x}
\geq 0
\label{j07}
\end{align}
holds in the second term of the left-hand side of \eqref{j02}. The above inequality implies that a reverse power flow occurs on the feeder line. 

In addition to the voltage subsystem, considering the equality~\eqref{j04} for the phase subsystem $\mathsf{P}$, the gradient of the voltage phase  at the  transformer can take both positive and negative values depending on the active and reactive powers from the following equality: 
\begin{align*}
\frac{{\rm d}\theta(0)}{{\rm d}x}=\int_0^L \frac{bp(x)-gq(x)}{g^2+b^2}{\rm d}x.
\end{align*}
Combining the assumption $q(x) \geq 0$ with the above equality, we see 
that the active and reactive powers satisfy the inequality 
\begin{align*}
0 \leq \int_0^Lq(x){\rm d}x \leq \frac{b}{g}\int_0^Lp(x){\rm d}x
\end{align*}
if and only if the following inequality holds:
\begin{align} 
\frac{{\rm d}\theta(0)}{{\rm d}x} \geq 0.
\label{j06}
\end{align}
The above inequality shows an occurrence of a phase advance on the feeder line, 
which often occurs in current power distribution systems including PVs and EVs.

We can see that the above discussion is consistent with typical phenomena observed power distribution systems including PVs and EVs. Therefore, they also show a validity of the main result given in Theorem~\ref{tm2}. 

\subsection{Discussion on reduction of distribution losses}

In this subsection, we discuss a reduction of the dissipation loss based 
on the main result. 

From \eqref{j02}, the integral of the dissipation rate  $\Delta(v(x),\theta(x))$ can be expressed as 
\begin{align}
\hspace*{-2mm}
\int_0^L\Delta(v(x),\theta(x)){\rm d}x
& = \int_0^L\frac{gp(x)+bq(x)}{g^2+b^2}{\rm d}x -  \frac{{\rm d}v(0)}{{\rm d}x}.
\label{j07}
\end{align}
We have shown that the distribution loss is determined by the dissipation rate $\Delta(v(x),\theta(x))$ of the voltage subsystem $\mathsf{V}$ in the previous subsection. 
The left-hand side of the above equality 
corresponds to the net distribution loss of the entire distribution system.
Thus, the equality~\eqref{j07} shows that the net loss is characterized by 
the supply rate $\frac{gp(x)+bq(x)}{g^2+b^2}$ of the voltage 
subsystem $\mathsf{V}$ and the voltage gradient $\frac{{\rm 
d}v(0)}{{\rm d}x}$ at the transformer ($x=0$). In more detail, to reduce the net distribution loss, it is important to attenuate the effect of the voltage gradients caused by transformers by an appropriate injection of both active power and reactive powers. Such a theoretical discussion for reducing overall distribution losses has not been considered so far in the framework of the nonlinear ODE model. Therefore, the discussion given here can be considered as one of the contributions of this paper. It is desired to derive appropriate design methods for both active and reactive power in our future work.

\section{Conclusion}

In this paper, we have shown dissipativity of the subsystems corresponding to the active and reactive powers of the power distribution system consisting of a straight feeder line. As a main result, we have proved the dissipativity of the voltage and phase subsystems.  We have also provided a physical interpretation of the main results based on the dissipation equalities of these subsystems. Finally, we have given a discussion for a reduction of distribution losses based on the dissipation rate of the voltage subsystem. 

One of the future work is the voltage control of the actual distribution system, which also takes into account the distribution losses due to active and reactive power injections. In addition, it is desired to generalize the results of this paper to the case where the bifurcations are contained in the feeder line. 

\appendix 
\section{Dissipation Theory for Spatial One-Dimensional Systems}
\label{sec:prel}

In this appendix, we provide a preliminary notion of the dissipation theory~(\cite{wil:leas1}, \cite{wil:diss1}, \cite{wil:diss2}) to consider dissipativity of distribution systems described by the nonlinear ODE models in this paper.Specifically, we integrate the dissipation theory~(See \cite{pil:loss1}) for linear $n$-dimensional systems with both time and space as independent variables, and nonlinear one-dimensional systems with time as an independent variable~(See \cite{kha:nonl1}).Based on this integration, we extend the framework to nonlinear one-dimensional systems with independent variables, e.g. space, that are not necessarily causal such as time, and are contained in a finite interval.



Consider a nonlinear system $\mathsf{S}$ with the independent variable $x \in [0,L]$, where $[0,L] \subset \mathbb{R}$ is the finite interval to which $x$ belongs. We suppose that $x$ is an independent variable, e.g. space, which is not necessarily causal such as time. Then, $\mathsf{S}$ is described by the following nonlinear ODE: 
\begin{align}
\frac{{\rm d}z(x)}{{\rm d}x}=f(z(x),u(x))
\label{q01} 
\end{align}
where $z(x) \in \R^n$ and $u(x)\in\R^m$ are the state and input of this 
system. In \eqref{q01}, $f:\R^{n+m} \rightarrow \R^{n}$ is a nonlinear function.

We give the definition of the nonlinear systems considered in this paper as follows. 
\begin{definition}
\label{df3}
\textit{}
{\rm Consider a nonlinear system $\mathsf{S}$ in \eqref{q01} with the independent variable $x \in [0,L]$, where $[0,L] \subset \mathbb{R}$ is the finite interval to which $x$ belongs. Let $\Phi(z(x),u(x))$ be the power delivered to the nonlinear system $\mathsf{S}$ in \eqref{q01}. We call $\Phi(z(x),u(x))$ \emph{supply rate of the system $\mathsf{S}$}. 
\begin{itemize}
\item[(i)]
The system $\mathsf{S}$ is called \emph{dissipative with respect to the supply rate} $\Phi(z(x),u(x))$ 
if there exist functions $\Psi: \mathbb{R}^n \rightarrow \mathbb{R}$ and $\Delta: \mathbb{R}^{n+m} 
\rightarrow \mathbb{R}$ satisfying the 
inequality
\begin{equation}
\frac{{\rm d}\Psi(z(x))}{{\rm d} x}  = \Phi(z(x),u(x)) - \Delta(z(x),u(x))
 \label{q02}
\end{equation}
and $\Delta(z(x),u(x)) \geq 0$ for any $x \in [0,L]$. 
\item[(ii)]
The system $\mathsf{S}$ is 
called \emph{dissipative with respect to the supply rate} $\Phi(z(x),u(x))$ 
if there exists a function $\Psi: \mathbb{R}^n \rightarrow \mathbb{R}$ satisfying the 
inequality 
\begin{equation}
\frac{{\rm d}\Psi(z(x))}{{\rm d} x}  = \Phi(z(x),u(x))
 \label{q03}
\end{equation}
 for any $x \in [0,L]$. 
\end{itemize}
In \eqref{q02} and \eqref{q03}, the functions $\Psi(z(x))$ and $\Delta(z(x),u(x))$ are called \emph{flux} and \emph{dissipation rate for $\mathsf{S}$ with respect to the supply rate $\Phi(z(x),u(x))$}. Moreover, the equalities~\eqref{q02}  and \eqref{q03} are called the \emph{dissipation equality}.}
\end{definition}

We can regard $\Phi(z(x),u(x))$ as the power delivered to the system $\mathsf{S}$. Moreover, the flux $\Psi(z(x))$ corresponds to the energy which moves over the interval $[0,L]$. From this point, the dissipation equality~\eqref{q02} implies that the spatial variation of the energy of the system does not exceed the power supplied to the system due to the existence of the dissipation rate which is a nonnegative function. This implies that the system dissipates energy to the external environment of the system. Note that we do not introduce time as an independent variable in this paper. For this reason, any storage function do not appear in this paper, which expresses the internal energy in the ordinary dissipation theory (\cite{pil:loss1}, \cite{wil:leas1} and etc.).

By computing the integral of the dissipation equality~\eqref{e02} from 
$x=0$ to $x=L$, we have the inequality 
\begin{align*}
\int_{0}^L & \Phi(z(x),u(x)) {\rm d}x - \left(\Psi(z(L)) + \Psi(z(0))\right)\\
= & \int_{0}^L \Delta(z(x),u(x)) {\rm d}x\\
\geq & 0.
\end{align*}
If we consider the space as the interval of infinite length, it is known that the dissipativity of the system is guaranteed only by the integral of the supply rate, i.e. the net power supplied to the system (See \cite{pil:loss1}). However, the space is restricted to an interval of finite length in this paper. Therefore, the compensation by the spatial variation of the flux results in dissipation of the energy of the entire system.

\end{document}